\documentclass{intech}

\usepackage{amsmath,amssymb,amsbsy,amsfonts}

\usepackage{graphicx}

\usepackage{natbib}

\booktitle{Will-be-set-by-IN-TECH}%

\chaptertitle{Extending cosmology: the metric 
              approach} 

\authors{Sergio Mendoza}
\affiliation{Instituto de Astronomia, Universidad Nacional Aut\'onoma de
Mexico, AP 70-264, Ciudad Universitaria, Distrito Federal 04510}
\country{Mexico}

\begin{document}

\maketitle

\section{Introduction}
\label{introduction}

  In this chapter it is shown how the introduction of a fundamental
constant of nature with dimensions of acceleration into the theory of
gravity makes it possible to extend gravity in a very consistent manner.
In the non-relativistic regime a MOND-like theory with a modification
in the force sector is obtained.  This description turns out to be
the the weak-field limit of a more general metric relativistic theory
of gravity.  The mass and length scales involved in the dynamics of
the whole universe require small accelerations which are of the order
of Milgrom's acceleration constant, it turns out that this relativistic
theory of gravity can be used to explain the expansion of the universe.
In this work it is explained how to build that relativistic theory
of gravity in such a way that the overall large-scale dynamics of the
universe can be treated in a pure metric approach without the need to
introduce dark matter and/or dark energy components.

  Cosmological and astrophysical observations are generally
explained introducing two unknown mysterious dark components, namely
dark matter and dark energy.  These ad hoc hypothesis represent a big
cosmological paradigm, since they arise due to the fact that Einstein's
field equations are forced to remain unchanged under certain observed
astrophysical phenomenology.

  A natural alternative scenario would be to see whether viable
cosmological solutions can be found if dark unknown entities are
assumed non-existent.  The price to pay with this assumption is that
the field equations of the theory of gravity need to be extended and so,
new Friedmann-like equations will arise.  The most natural approach to
extend gravity arises when a metric extension \( f(R) \) is introduced
into the theory \citep[see e.g.][and references therein]{capozziellobook}.

  In a series of recent articles,
\citet{carranza12,mendoza11,bernal11,hernandez10,hernandez12a,bernal11a,mendoza12}
have shown how relevant the introduction of a new fundamental physical
constant \( \mathsf{a}_0 \approx 10^{-10} \textrm{m}/\textrm{s}^2
\) with dimensions of acceleration is in excellent agreement with
different phenomenology at many astrophysical mass and length
sizes, from solar-system to extragalactic and cosmological scales.
The introduction of the so called Milgrom's acceleration constant \(
\mathsf{a}_0 \) in a description of gravity means that any gravitational
field produced by a certain distribution of mass (and hence energy)
needs to incorporate the acceleration \( \mathsf{a}_0 \) together with
Newton's gravitational constant \( G \) and the speed of light \( c \)
in the description of gravity.

  In section~\ref{extended-newtonian} it is shown, through
a description of an extended Newtonian gravity scenario, the
advantages of working with a modification of gravity dependent on
the mass and lengths associated with the dimensions and masses
of the sources that generate  the gravitational field, and not
with the dynamical acceleration they produce on test particles.
Section~\ref{relativistic-extension} describes how it is possible
to build a metric theory of gravity which generalises the extended
Newtonian description mentioned in section~\ref{extended-newtonian} and
section~\ref{frt-connection}  interconnects this extended relativistic
description of gravity with a metric description of gravity for which
the energy-momentum tensor appears in the gravitational field's action.
On section~\ref{cosmological-applications} we use the developed theory of
gravity for cosmological applications in a dust universe and see how it is
a coherent representation of gravity at cosmological scales.  Finally on
section~\ref{discussion}, we discuss the consequences of the developed
approach of gravity and some of the future developments of the theory.

\section{Extended Newtonian gravity}
\label{extended-newtonian}

  \citet{milgrom83c,milgrom08-paradigm,milgrom10}  constructed a
MOdified Newtonian Dynamics (MOND) theory, based on the introduction of a
fundamental constant of nature \( \mathsf{a}_0 = 1.2 \times 10^{-10} \textrm{m}
\,\textrm{s}^{-2} \) in such a way that the acceleration experienced by a
test particle on a gravitational field produced by a point mass source \( M
\) is such that:

\begin{equation}
  a = \left\{
      \begin{array}{c l}
        &-  \frac{ GM }{r^2}, \qquad \qquad \text{for} \qquad \qquad a 
	\gg \mathsf{a}_0, \\
        &-  \frac{\sqrt{\mathsf{a}_0 G M }}{r}, \qquad \text{\ \ for} \qquad 
	\qquad  a \ll \mathsf{a}_0,
     \end{array}
      \right.
\label{eq001}
\end{equation}

\noindent where \( r \) is the radial distance to the central mass.
In other words, for accelerations \( a \gg \mathsf{a}_0 \), Newtonian
gravity is recovered and new MONDian effects are expected to appear
for accelerations \(  a \lesssim \mathsf{a}_0 \).   The strong \( a \ll
\mathsf{a}_0 \) MONDian regime means that Kepler's third law is not valid
since for a circular orbit about the central mass \( M \), the acceleration
\( a = v / r \), where \( v \) is velocity of the test mass, and so \(
v = \left( \mathsf{a}_0 G M \right)^{1/4} \propto M^{1/4} \), which
is the Tully-Fisher relation \citep[see e.g.][]{puech10} for the 
case of a spiral galaxy and is the same relation experienced by wide-open
binaries \citep{hernandez12a} and by the tail of the ``rotation curve'' in
globular clusters \citep{hernandez12,hernandez12b}.

  In order to interpolate from the strong \( a \gg \mathsf{a}_0 \)
Newtonian regime to the weak \( a \ll \mathsf{a}_0 \) one, the traditional
MONDian approach is to construct a somewhat built-by-hand interpolation
function  \( \mu(y) \)  in such a way that

\begin{gather}
  a \mu(y) = - \frac{ G M }{ r^2 },
			\label{eq002} \\
  \intertext{where} 
  \mu(y)  = \begin{cases}
              1, \qquad \text{for} \qquad y \gg 1,\\
	      y, \qquad \text{for} \qquad y \ll 1,
            \end{cases}
  \text{and} \quad y:= \frac{ a }{ \mathsf{a}_0 }.
  					\notag
\end{gather}

  The usual approach to MOND as expressed by equation~\eqref{eq002}
means that Newton's 2nd law of mechanics needs to be modified
\citep[see e.g.][]{bekenstein06b}.  As explained by \citet{mendoza11}, a
better physical approach can be constructed if the modification is made in
the force (gravitational) sector.  Indeed, by the use of Buckingham's
theorem of dimensional analysis \citep[cf.][]{sedov},  the gravitational
acceleration experienced by a test particle is given by

\begin{gather}
  a = \mathsf{a}_0 g(x),  
  			\label{eq003}\\
  \intertext{where the dimensionless quantity}
    x := \frac{ l_M }{  r  },
      			\label{eq004} \\
  \intertext{and a mass-length scale}
  l_M := \left( \frac{ G M }{\mathsf{a}_0} \right)^{1/2}.
  			\label{eq005}
\end{gather}

  The length \( l_M \) plays an important role in the description of the
theory and is such that when \( l_M \gg r \), the strong Newtonian regime
of gravity is recovered and when \( l_M \ll r \) the weak MONDian regime of
gravity appears.  As such, the dimensionless acceleration (or \emph{transition
function}) \( g(x) \) is such that:

\begin{equation}
  \frac{ a }{ \mathsf{a}_0 } = g(x) :=
    \begin{cases}
      x^2, \qquad \text{when} \quad x \gg 1,  \\
      x,   \qquad \text{ when} \quad x \ll 1.
    \end{cases}
\label{eq006}
\end{equation}

  In general terms, a mass distribution whose
length is much greater than its associated mass-length \( l_\text{M} \)
is in the MONDian regime (since \( x \ll 1 \)) and a mass distribution
whose length is much smaller than its mass-length scale is in the
Newtonian regime (since \( x \gg 1 \)). The case \( x = 1 \) can roughly
be thought of as the point where the transition from the Newtonian to
the MONDian regime occurs.

  A general transition function \( g(x) \) was
built by \cite{mendoza11} taking  Taylor expansion series
about the correct MONDian and Newtonian limits, yielding:

\begin{equation}
  g(x) = x \, \frac{ 1 \pm x^{n+1} }{ 1 \pm x^n
    }.
\label{eq007}
\end{equation}

\begin{figure}
\begin{center}
  \includegraphics[scale=0.7]{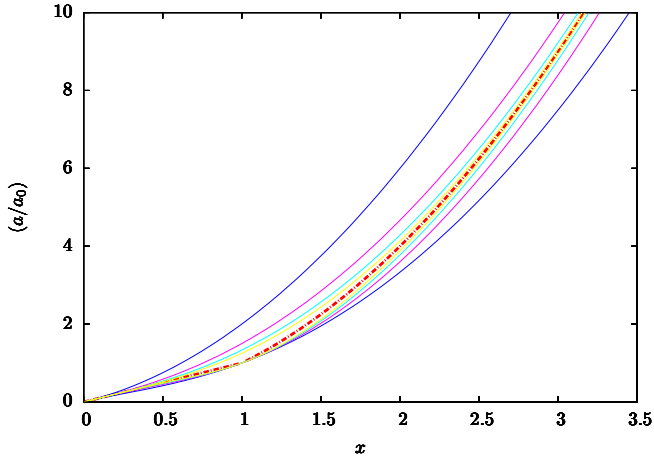}
\end{center}
\caption[Acceleration]{ The figure taken from
  \citet{mendoza11} shows the acceleration function \( a \) in units
  of Milgrom's acceleration constant \( \mathsf{a}_0 \) as a function
  of the parameter \( x \). The thick dash-dot curve is the extreme
  limiting value \( n \to \infty \), i.e. \( a / \mathsf{a}_0 = x \) for
  \( x \leq 1 \) and \( a / \mathsf{a}_0 = x^2 \) for \( x \geq 1 \).
  The curves above and below this extreme acceleration line represent
  values of \( n = 4,\ 3,\ 2,\ 1, \) for the minus and plus signs of
  equation \eqref{eq007} respectively.  The extreme limiting curve has
  a kink at \( x = 1 \).  }
\label{fig01}
\end{figure}

\noindent This non-singular function converges to the correct expected
limits of equation~\eqref{eq006} for any value of the parameter \( n
\geq 0 \).  As shown in Figure~\ref{fig01}, the transition function \(
g(x) \) rapidly converges to the limit ``step function''

\begin{equation}
  g(x)\bigg|_{n\rightarrow\infty}  = 
  \begin{cases}
    x, \qquad \text{  for }  0 \leq x \leq 1, \\
    x^2, \qquad \text{for  } x \geq 1,
  \end{cases}
\label{eq008}
\end{equation}

\noindent when \( n \gtrsim 3 \).  The parameter \( n \) needs to be
found empirically by astronomical observations.  The value found by
\citet{mendoza11} for the rotation curve of our galaxy is \( n \gtrsim
3 \) and the one found by \citet{hernandez12,hernandez12a,hernandez12b}
is  \( n \gtrsim 8 \), with a minus sign selection on the numerator
and denominator on the right hand side of equation~\eqref{eq007}.   These
authors have shown that a large value of \( n \) is coherent with solar
system motion of planets, rotation curves of spiral galaxies, equilibrium
relations of dwarf spheroidal galaxies and their correspondent relations
in globular clusters, the Faber-Jackson relation and the fundamental plane
of elliptical galaxies as well as with the orbits of wide binary stars.
The \( n =3 \) model in which a small, but measurable transition is
obtained, has also been tested on earth and moon-like experiments by
\citet{meyer11}  and \citet{exirifard11} respectively, showing that it
is coherent with such precise measurements.  In  fact, these experiments
also validate all \( n \geq 3 \) models.

  Care must be taken when the introduction of a new fundamental constant
of nature with dimensions of acceleration \( \mathsf{a}_0 \) is made.
In fact, the introduction of \( \mathsf{a}_0 \) does not impose any kind
causality arguments such as the ones given by the velocity of light \(
c \).  In fact, one may think of \( \mathsf{a}_0 \) as a fundamental
constant needed to transit from one gravity regime to another.  In this
respect for example, instead of using \( \mathsf{a}_0  \) as a fundamental
constant, one may define

\begin{equation}
  \Sigma_0 := \mathsf{a}_0 / G =  1.8 \, \mathrm{kg} \, \mathrm{m}^{-2},
\label{eq40}
\end{equation}

\noindent as the new fundamental constant of nature.  The constant \(
\Sigma_0 \), with dimensions of surface mass density, enters in the
description of the gravitational theory in such a way that
equations~\eqref{eq003} and \eqref{eq005} are given by:

\begin{gather}
  a = - G \Sigma_0 g\left(l_M/r\right), \qquad 
    l_M := \left( M / \Sigma_0 \right).
				\label{eq41} \\
\intertext{and the acceleration in the full MONDian regime and the
corresponding Tully-Fisher relation are}
  a = - G \frac{ \left( \Sigma_0 M \right)^{1/2} }{ r }, \qquad 
    v =  G^{1/2} \Sigma_0^{1/4} M^{1/4}.
				\label{eq42}
\end{gather}

\noindent Also, a more manageable extended fundamental quantity, directly
measurable through the Tully-Fisher relation, can be defined:

\begin{equation}
  \epsilon_M := \mathsf{a}_0 G = 8.004 \times 10^{-21} \mathrm{m}^4 \,
    \mathrm{s}^{-4} \, \mathrm{kg}^{-1},
\label{eq43}
\end{equation}
  
\noindent with dimensions of velocity to the fourth over mass, for which

\begin{gather}
  a = - \frac{ \epsilon_M }{ G } g\left(l_M/r\right), \qquad 
    l_M := \left( G^2 \, M / \epsilon_0 \right).
				\label{eq44} \\
\intertext{With this, the acceleration of a test particle in the full 
MOND regime and the Tully-Fisher relation are:}
  a = - \frac{ \left( \epsilon_M \, M \right)^{1/2} }{ r }, \qquad 
    v =  \left( \epsilon_M \, M \right)^{1/4}.
				\label{eq45}
\end{gather}

  The choice of a new fundamental constant of nature has many ways in which
it can be introduced into the theory \citep{sedov}.  In this work, the use
of \( \mathsf{a}_0 \) is kept as it is traditionally done, but we note the
fact that \( \epsilon_0 \) is the best fundamental constant to use since it
is directly measured through the flattened rotation curves of spiral
galaxies.

  The extended Newtonian model of gravity presented in this section
is equivalent with MOND on spherical and cylindrical symmetry but
deviates considerable from it for systems away from this symmetry
\citep{mendoza11}.  As we have already shown, there are however many
advantages of using this approach, the most objective meaning that
the modification is made on the force sector and not a modification on
the dynamics.

\section{Relativistic metric extension}
\label{relativistic-extension}

  Finding a relativistic theory of gravity for which one
of its non-relativistic limits converges to MOND yields
usually strange assumptions and/or complicated ideas \citep[see
e.g.][]{mishra12,blanchet12,bekenstein04}.  A good first approach was
provided by a slight modification of Einstein's field equations by
\citet{sobouti06}, but the attempt is not complete.

  In order to find an elegant and simple theory of gravity for which
a MONDian solution is found, \citet{bernal11} used a metric correct
dimensional interpretation of Hilbert's gravitational action \( S_\text{f}
\) in such a way that:

\begin{equation}
   S_\text{f}  = - \frac{ c^3 }{ 16 \pi G L_M^2 } \int{ f(\chi) \sqrt{-g}
     \, \mathrm{d}^4x},
\label{eq010}
\end{equation}

\noindent which slightly differs from its traditional form (see
e.g. \cite{capozziellobook,sotiriou10,capozziello10a}) since the following
dimensionless quantity has been introduced:

\begin{equation}
  \chi := L_M^2 R,
\label{eq011}
\end{equation}

\noindent where \( R \) is Ricci's scalar and \( L_M \) defines a length
fixed by the parameters of the theory.  The explicit form of the length
\( L \) has to be obtained once a certain known limit of the theory is
taken, usually a non-relativistic limit.  Note that the definition of \(
\chi \) gives a correct dimensional character to the action~\eqref{eq010},
something that is not completely clear in all previous works dealing
with a metric description of the gravitational field.  For \( f(\chi)
= \chi \) the standard Einstein-Hilbert action of general relativity 
is obtained.

  On the other hand, the matter action has its usual form,

\begin{equation}
  S_\text{m} = - \frac{ 1 }{ 2 c } \int{ {\cal L}_\text{m} \, \sqrt{-g} \,
    \mathrm{d}^4x },
\label{eq012}
\end{equation}

\noindent with \( {\cal L}_\text{m} \) the matter Lagrangian density of the
system.  The null variations of the complete action, i.e. \( \delta
\left( S_\text{H} + S_\text{m} \right) = 0 \), yield the following
field equations:

\begin{equation}
  \begin{split}
    f'(\chi) \, \chi_{\mu\nu} - \frac{ 1 }{ 2 } f(\chi) g_{\mu\nu} - L_M^2 &
      \left( \nabla_\mu \nabla_\nu -g_{\mu\nu} \Delta \right) f'(\chi)
  				\\
    &= \frac{ 8 \pi G L_M^2 }{ c^4} T_{\mu\nu},
  \end{split}
\label{eq013}
\end{equation}

\noindent where the dimensionless Ricci tensor \( \chi_{\mu\nu} \) 
is given by:

\begin{equation}
  \chi_{\mu\nu} := L_M^2 R_{\mu\nu},
\label{eq014}
\end{equation}

\noindent and \( R _{\mu\nu} \) is the standard Ricci tensor.
The Laplace-Beltrami operator has been written as \( \Delta :=
\nabla^\alpha \nabla_\alpha \) and the prime denotes derivative with
respect to its argument.  The energy-momentum tensor \( T_{\mu\nu} \)
is defined through the following standard relation: \( \delta S_\text{m}
= - \left( 1 / 2 c \right) T_{\alpha\beta} \, \delta g^{\alpha\beta} \).
In here and in what follows, we choose a (\(+,-,-,-\)) signature for
the metric \( g_{\mu\nu} \) and use Einstein's summation convention over
repeated indices.

  The trace of equation~\eqref{eq013} is:
\begin{equation}
  f'(\chi) \, \chi  - 2 f(\chi) + 3 L_M^2  \, \Delta  f'(\chi) = 
    \frac{ 8 \pi G L_M^2 }{ c^4} T,
\label{eq015}
\end{equation}

\noindent where \( T := T^\alpha_\alpha \).

  In order to search for a MONDian solution, \citet{bernal11}
analysed the problem in two ways.  First by performing an order of
magnitude approach to the problem, and second, by doing a full perturbation
analysis.  Since the second technique is merely to fix constants of
proportionality of the problem, their order of magnitude approach and its
consequences are discussed in the remain of this section.
Also, since we are interested at the moment on a point mass distribution 
generating a stationary spherically symmetric
space-time, the trace equation~\eqref{eq015} contains all the
relevant information relating the field equations.
At this point it is also useful to assume a power~law form for the 
function 

\begin{equation}
  f(\chi) = \chi^b.
\label{fchi-power}
\end{equation}

\noindent An order of magnitude approach to the
problem  means that \( \mathrm{d} / \mathrm{d} \chi \approx 1 / \chi \), \(
\Delta \approx - 1 / r^2  \) and the mass density \( \rho \approx  M /
r^3 \).  With this, the trace~\eqref{eq015} takes the following form:

\begin{equation}
   \chi^b  \left( b - 2 \right) - 3 b L_M^2  \frac{ \chi^{(b-1)} }{ r^2 }
     \approx \frac{ 8 \pi G M L_M^2 }{ c^2 r^3}.
\label{eq017}
\end{equation}
  
  Note that the second term on the left-hand side of equation~\eqref{eq017}
is much greater than the first term when the following condition is
satisfied:

\begin{equation}
  R  r^2  \lesssim \frac{ 3 b }{ 2 - b }.
\label{eq018}
\end{equation}

\noindent  At the same order of approximation, Ricci's scalar \( R
\approx \kappa = R_\text{c}^{-2} \), where \( \kappa \) is the Gaussian
curvature of space and \( R_\text{c} \) its radius of curvature and so,
relation~\eqref{eq018} essentially means that

\begin{equation}
  R_c \gg r.
\label{eq019}
\end{equation}

\noindent  In other words, the second term on the left-hand side of
equation~\eqref{eq017} dominates the first one when the local radius of
curvature of space is much grater than the characteristic length \(
r \).  This should occur in the weak-field regime, where MONDian effects
are expected.  For a metric description of gravity, this limit must
correspond to the relativistic regime of MOND.

  Under assumption~\eqref{eq019}, equation~\eqref{eq017} takes the following
form:

\begin{equation}
    R^{ (b-1)} \approx - \frac{ 8 \pi G M  }{ 3 b c^2 r
      L_M^{2 \left( b - 1 \right) } }.
\label{eq020}
\end{equation}

  We now recall the well known relation followed by the Ricci scalar at
second order of approximation at the non-relativistic level 
\cite{daufields}:

\begin{equation}
  R =  - \frac{ 2 }{  c^2 } \nabla^2 \phi = +\frac{ 2 }{ c^2 } \nabla \cdot
    \boldsymbol{a},
\label{eq021}
\end{equation}

\noindent where the negative gradients of the gravitational potential \(
\phi \) provide the acceleration \( \boldsymbol{a} := - \nabla \phi \)
felt by a test particle on a non-relativistic gravitational field. At
order of magnitude, equation~\eqref{eq021} can be approximated as

\begin{equation}
  R \approx - \frac{ 2 \phi }{  c^2 r^2 } \approx  \frac{ 2 a }{  c^2 r }.
\label{eq022} 
\end{equation}

  Substitution of this last equation on relation~\eqref{eq020} gives

\begin{eqnarray}
  a &\approx&   - \frac{ c^2 r }{ 2 L_M^2 }  \left( \frac{ 8 \pi G M  }{
    3 b c^2 r  } \right)^{1/\left( b - 1 \right)},\nonumber\\ & \approx
    & - c^{\left( 2 b- 4  \right)/\left( b - 1 \right) } r^{ \left( b -
    2 \right) / \left( b - 1 \right) } L_M^{-2} \left( G M  \right)^{
    1 / \left( b - 1 \right) }.
\label{eq023}
\end{eqnarray}

  This last equation converges to a  MOND-like acceleration \( a \propto
1 / r \) if \( b - 2 = - \left( b - 1  \right) \), i.e. when  \( b = 3 / 2
\). Also, at the lowest order of approximation, in the extreme
non-relativistic limit, the velocity of light \( c \) should not appear
on equation~\eqref{eq023} and so, the only way this condition is fulfilled
is that \( L_M \)
depends on a power of \( c \), i.e.

\begin{equation}
  L_M^{ - 2 } \propto c^{\left( 4 - 2 b \right)/\left( b - 1 \right) }  =
    c^2, \quad \text{and so,} \qquad L_M \propto c^{-1}.
\label{eq021a}
\end{equation}

  As discussed by \citet{bernal11}, the length \( L_M \) must
be constructed by fundamental parameters describing the theory of 
gravity and since the only two characteristic lengths of the problem are
the mass-length \( l_M \) and the gravitational radius 

\begin{equation}
  r_\text{g} = \frac{ G M }{ c^2 },
\label{eq022a}
\end{equation}

\noindent then the correct dimensional form of the length \( L_M \) is
given by 

\begin{equation}
  L_M = \zeta \, r_\text{g}^\alpha l_M^\beta, \qquad \text{with} \qquad
    \alpha + \beta = 1,
\label{eqlm}
\end{equation}

\noindent where the constant of proportionality \( \zeta \) is a
dimensionless number that can be found by a full perturbation
analysis technique and is given by \citep{bernal11}:

\begin{equation}
  \zeta = \frac{ 2 \sqrt{2} }{ 9 },
\label{eqzeta}
\end{equation}

  Substituting equation~\eqref{eqlm}  and the
value \( b = 3/2 \)  into relation~\eqref{eq021a}, it then follows that

\begin{equation}
  \alpha = \beta = 1/2, \qquad \text{i.e.} \qquad L_M \approx
    r_\text{g}^{1/2} l_M^{1/2} \, .
\label{eq22}
\end{equation}

  If we now substitute this last result and the value \( b = 3/2 \) 
in equation~\eqref{eq023}  we get:

\begin{equation}
  a \approx - \frac{ \left( \mathsf{a}_0 G M \right)^{1/2} }{ r },
\label{eq023b}
\end{equation}

\noindent which is the traditional form of MOND for a point mass source
(see e.g.  \cite{milgrom09,milgrom10,bekenstein06c} and references
therein).  Also, the results of equation~\eqref{eq023b} in~\eqref{eq022}
mean that

\begin{equation}
  R \approx \frac{ r_\text{g} }{ l_M } \, \frac{ 1 }{ r^2 },
\label{eq030}
\end{equation}

\noindent and so, inequality~\eqref{eq019} is equivalent to

\begin{equation}
  l_M \gg r_\text{g}.
\label{eq031}
\end{equation}

\noindent  The regime imposed by equation~\eqref{eq031} is precisely
the one for which MONDian effects should appear in a relativistic theory
of gravity.  This is an expected generalisation of the results presented
in section~\ref{extended-newtonian}.  Note that in the weak field limit
regime for which \( l_M \ll r \) together with equation~\eqref{eq031}
yields \( r \gg l_M \gg r_\text{g} \).  In this connection, we also note
that Newton's theory of gravity is recovered in the limit \( l_M \gg r
\gg r_\text{g} \).

  In exactly the same way as it was done to build the transition function
for the case of extended Newtonian gravity in
section~\ref{extended-newtonian}, a general function \( f(\chi)  \) can 
be constructed:

\begin{equation}
   f(\chi) = 
      \chi^{3/2} \, \frac{ 1 \pm \chi^{p+1} }{ 1 \pm \chi^{3/2 + p
        } } \rightarrow
    \begin{cases}
      \chi^{3/2}, \quad \text{for } \chi \ll 1, \\
             \chi, \qquad \ \text{for } \chi \gg 1.  
    \end{cases}
\label{eq032}
\end{equation}

\begin{figure}
\begin{center}
  \includegraphics[scale=0.7]{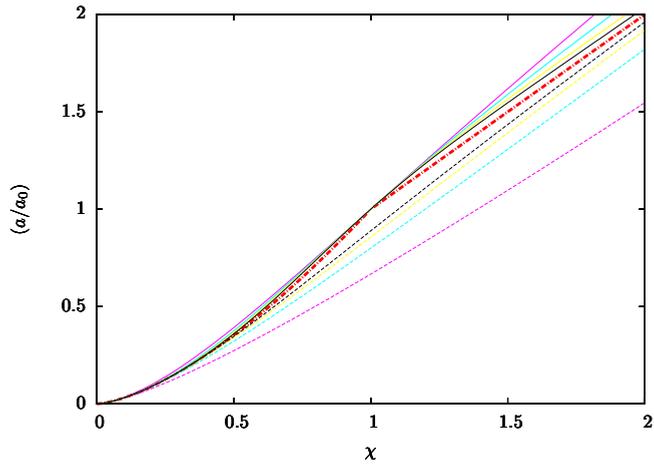}
\end{center}
\caption[Acceleration relativistic]{ The figure shows the
  transition function \( f(\chi) \), as a function of the dimensionless
  Ricci scalar \( \chi \),  for different regimes of gravity, converging
  to \( f(\chi) = \chi \) for \( \chi \gg 1 \) (general relativity) and
  to \( f(\chi) = \chi^{3/2} \) for \( \chi \ll 1 \) (a relativistic
  regime with MOND as its weak field limit) -see equation~\eqref{eq032}.
  The thick dash-dot curve is the extreme limiting value \( p \to \infty
  \), i.e. \( f(\chi) = \chi^{3/2} \) for \( \chi \leq 1 \) and \(
  f(\chi) = \chi \) for \( \chi \geq 1 \).  The curves above and below
  this extreme  function represent values of \( p = 3,\ 2,\ 1, \ 0\)
  for the minus and plus signs of equation \eqref{eq007} respectively.
  The extreme limiting curve has a kink at \( \chi = 1 \).  }
\label{fig03}
\end{figure}

\noindent In other words, general relativity is recovered when \( \chi
\gg 1 \) in the strong field regime and the relativistic version of MOND
with \( \chi^{3/2} \) is recovered for the weak field regime of gravity
when \( \chi \ll 1 \) (see Figure~\ref{fig03}). The unknown parameter \(
p \geq -1  \)  needs to be calibrated with astronomical observations,
in an analogous form  as the calibration of the parameter \( n \)
in equation~\eqref{eq007} was done.  This is a much harder task and a
matter of future research. However, since the non-relativistic approach
to gravity explained in section~\ref{extended-newtonian} means that the
transition from the Newtonian to the MONDian regimes of gravity is very
sharp, it  most probably means that the function \( f(\chi)  = \chi \)
for \( \chi \geq 1 \) and that \( f(\chi) = \chi^{3/2} \) for \( \chi
\leq 1 \), but this has to be tested by some astronomical observations.

  The mass dependence of \( \chi \) and \( L_M \) mean that Hilbert's
action~\eqref{eq010} is a function of the mass \( M \).  This is usually not
assumed, since that action is thought to be purely a function of the
geometry of space-time due to the presence of mass and energy sources.
However, it was Sobouti \cite{sobouti06} who first encountered
this peculiarity in the  Hilbert action when dealing with a metric
generalisation of MOND.  Following the remarks by \citet{sobouti06} and
\citet{mendoza07} one should not be surprised if some of the commonly
accepted notions, even at the fundamental level of the action, require
generalisations and re-thinking.  An extended metric theory of gravity
goes beyond the traditional general relativity ideas and in this way,
we need to change our standard view of its fundamental principles.

\section{$F(R,T)$ connection}
\label{frt-connection}

  For the description of gravity shown in
section~\ref{relativistic-extension} it follows that an adequate way of
writing up the gravitational field's action is given by:

\begin{equation}
   S_\text{f}  = - \frac{ c^3 }{ 16 \pi G  } \int{ \frac{ f(\chi) }{ L_M^2
     } \sqrt{-g} \, \mathrm{d}^4x}.
\label{eq040}
\end{equation}

\noindent The function \( L_M \) is a function of the mass of the system
and in general terms it is a function of the space-time coordinates.
For the particular case of a spherically symmetric space-time it coincides
with the mass of the central object generating the gravitational field as
expressed in equations~\eqref{eqlm} and~\eqref{eq22}.  Generally speaking
what the meaning of \( M \) would be for a particular distribution
of mass and energy needs further research, beyond the scope of
this work.  Nevertheless one expects that for dust systems with spherically
symmetric distributions, the function \(
M \) would be given by the standard mass-energy relation \citep[see
e.g.][]{MTW}:

\begin{equation}
  M := \frac{ 4 \pi }{ c^2 } \int{ T \, r^2 \, \mathrm{d}r },
\label{eq041}
\end{equation}

 In very general terms, the definition of \( M \) in this last equation
means that \( M \) would not be invariant.  However, in some particular
systems with high degree of symmetry it is possible to make this quantity
invariant.  For example, in the case of a spherically symmetric spacetime
produced by a point mass that quantity is simply the ``Schwarzschild''
mass of the point mass generating the gravitational field.  In the
cosmological case it is also possible to define it as an invariant
quantity as discussed in section~\ref{cosmological-applications}.

  The field equations produced by the null variations of the addition of
the field's action \( S_\text{f} + S_\text{m} \) can be constructed in the
following form.  \citet{harko11} have built an \( F(R,T) \) theory of
gravity, so making the natural identification:

\begin{equation}
  F(R,T) := \frac{ f(\chi)  }{ L_M^2 },
\label{eq03}
\end{equation}

\noindent it is possible to use all their results for our particular case
expressed in equation~\eqref{eq03}.
For example,  the null variations of the complete action 
\( S_\text{f} + S_\text{m} \) for the particular case of
equation~\eqref{eq03} is given by \citet{harko11}:

\begin{equation}
  \begin{split}
    \left( \frac{ f_R }{ L_M^2 } \right) \, R_{\mu\nu} -& \frac{ 1 }{
      2 L_M^2 }  f \, g_{\mu\nu} + \bigg[ g_{\mu\nu} \Delta - \nabla_\mu
      \nabla_\nu \bigg] \left( \frac{ f_R }{ L_M^2 } \right) =
     							\\ 
    &\frac{ 8 \pi G }{ c^4 } T_{\mu\nu} - \left( \frac{ f }{ L_M^2 }
      \right)_T \bigg[ T_{\mu\nu} + \Theta_{\mu\nu} \bigg],
  \end{split}
\label{eq07}
\end{equation}

\noindent and its trace is given by:

\begin{equation}
  \frac{ f_R \, R }{  L_M^2 } - \frac{ 2 f }{ L_M^2 } + 3 \Delta \left(
    \frac{ f_R }{ L_M^2 } \right) = \frac{ 8 \pi G }{ c^4 } T - \left(
    \frac{ f }{ L_M^2 } \right)_T \bigg[ T + \Theta \bigg],
\label{eq08}
\end{equation}

\noindent where the subscripts \( R \) and \( T \) stand for the partial
derivatives with respect to those quantities, i.e.

\begin{equation}
  \bigg( \ \ \bigg)_R := \frac{ \partial }{ \partial R }, \qquad
    \text{and} \qquad
  \bigg( \ \ \bigg)_T := \frac{ \partial }{ \partial T}.
\label{eq08a}
\end{equation}

\noindent The tensor \( \Theta_{\mu\nu} \) is such that 
\( \Theta_{\mu\nu} \delta g^{\mu\nu} := g^{\alpha\beta} \delta
T_{\alpha\beta} \)  and for the case of an ideal fluid it can be written as
\citep{harko11}:

\begin{equation}
  \Theta_{\mu\nu} = - 2 T_{\mu\nu} - p g_{\mu\nu}.
\label{eq09}
\end{equation}

  Note that equation~\eqref{eq07} or~\eqref{eq08} converge to the
field~\eqref{eq013} and trace~\eqref{eq015} relations as discussed in
section~\ref{relativistic-extension} when one considers a point mass
generating the gravitational field, i.e. when \( L_M = \text{const.} \)
and so \( \partial / \partial R = L_M^2 \partial / \partial \chi \).

  In general terms, the \( F(R,T) \) theory described by \citet{harko11}
produces non-geodesic motion of test particles since:

\begin{equation}
  \begin{split}
  \nabla ^{\mu } & T_{\mu \nu }
     = \left( \frac{ f }{ L_M^2}\right)_{T}
      \left\{ \frac{ 8\pi G }{ c^4 } - \left( \frac{ f }{ L_M^2
      }\right)_{T} \right\}^{-1} \times \\
    & \left[ \left( T_{\mu \nu }+\Theta
      _{\mu \nu }\right) \nabla^{\mu } \ln \left( \frac{f}{L_M^2}
      \right)_{T}\left( R,T\right) + \nabla ^{\mu }\Theta _{\mu \nu
      }\right] \neq 0,
  \end{split}
\label{eq10}
\end{equation}

\noindent and as such the geodesic equation has a force term:

\begin{equation}
  \frac{d^{2}x^{\mu }}{ds^{2}}+\Gamma _{\nu \lambda }^{\mu }u^{\nu
    }u^{\lambda }=\lambda^{\mu },  
\label{eqmot}
\end{equation}

\noindent where the four-force 

\begin{equation}
  \begin{split}
    \lambda^{\mu }: =& \frac{ 8 \pi G }{ c^4 } \left( \rho c^2 + p
      \right)^{-1} \left[ \frac{ 8 \pi G }{ c^4 } 
      + \left( \frac{ f }{ L_M^2 } \right)_T \right]^{-1} \times \\
    & \left( g^{\mu \nu }
      - u^{\mu } u^{\nu } \right) \nabla _{\nu }p,
  \end{split}
\end{equation}

\noindent is perpendicular to the four velocity \( \mathrm{d}x^\alpha /
\mathrm{d} s \).  As explained by \citet{harko11}, the motion of test
particles is geodesic, i.e. \( \lambda^\mu = 0 \) and/or \( \nabla^\alpha
T_{\alpha\beta} = 0 \),  (i) for the case of a pressureless \( p = 0 \)
(dust) fluid and (ii) for the cases in which \( F_T(R,T) = 0 \).

  In what follows we will see how all the previous ideas can be applied
to a  Friedmann-Lema\^{\i}tre-Robertson-Walker dust universe and so,
the divergence of the energy momentum tensor in equation~\eqref{eq10}
is null.  It is worth noting that this condition on the energy-momentum
tensor for many applications needs to be zero, including applications
to the universe at any epoch.

\section{Cosmological applications}
\label{cosmological-applications}

  There are many good and interesting attempts to explain many
cosmological observations using modified theories of gravity
\citep[see e.g.][and references therein]{nojiri11}, however these
theories are not generally fully consistent with the gravitational
anomalies shown at galactic and extragalactic scales discussed in
sections~\ref{extended-newtonian} and~\ref{relativistic-extension}.
To see whether the gravitational \( f(\chi)\) theory developed
in the previous sections can deal with cosmological data, let us
now apply the results obtained in those sections to an isotropic
Friedmann-Lema\^{\i}tre-Robertson-Walker (FLRW) universe following
the procedures first explored by \citet{carranza12}.  In this case,
the interval \( \mathrm{d} s \) is given by \citep{galaxy-formation}:

\begin{equation}
  \mathrm{d}s^2 = c^2 \mathrm{d}t^2 - a^2(t) \left\{ \frac{ 
    \mathrm{d}r^2 }{ 1 - \kappa r^2 } + r^2 \mathrm{d}\Omega^2 \right\},
\label{eq11}
\end{equation}

\noindent where \( a(t) \) is the scale factor of the universe normalised
to unity, i.e. \( a_0 = 1 \), at the present epoch \( t_0 \), and 
the angular displacement \( \mathrm{d} \Omega^2 := \mathrm{d} 
\theta^2 + \sin^2\theta \, \mathrm{d} \varphi^2 \) for the polar 
\( \mathrm{d} \theta \) and azimuthal \( \mathrm{d} \varphi \) 
angular displacements with a comoving coordinate distance \( r \).  In what
follows we assume a null space curvature \( \kappa = 0 \) at the present
epoch in accordance with observations and deal with the expansion of the
universe dictated by the field equations~\eqref{eq07}, avoiding any form
of dark unknown component.  Since we are interested on the compatibility
of this cosmological model with SNIa observations, in what follows we
assume a dust \( p = 0 \) model for which the covariant divergence
of the energy-momentum tensor vanishes, and so as discussed in
section~\ref{frt-connection} the trajectories of test particles are
geodesic.

  To begin with, let us rewrite the field equations~\eqref{eq07} inspired
by the approach first introduced by \citet{capozziello02} (see also
\citet{capozziellobook}) as follows:

\begin{gather}
    G_{\mu\nu} = \frac{ 8 \pi G }{ c^4 } \left\{ \left( 1 + \frac{ c^4 }{ 
      8 \pi G } F_T \right) \frac{ T_{\mu\nu} }{  F_R } +
      T_{\mu\nu}^\text{curv} \right\},
					\label{eq12} \\
\intertext{where the Einstein tensor is given by its usual form:}
  G_{\mu\nu}:= R_{\mu\nu} - \frac{1}{2} R g_{\mu\nu}.
  					\label{eq13} \\
\intertext{and}
  \begin{split}
    T_{\mu\nu}^\text{curv} :=  \frac{ c^4 }{ 8\pi G F_R } 
      &\left[  \left(\frac{1}{2} \left( F - R F_R \right)
            - \Delta F_R \right) g_{\mu\nu} \right.  + \\
	    & \nabla_{\mu}\nabla_{\nu}F_R \bigg],
  \end{split}
					\label{eq14}
\end{gather} 

\noindent represents the ``\emph{energy-momentum}'' curvature tensor.  Since
\( T_{00} = \rho c^2 \),  then it will be useful the identification \(
T_{00} := \rho_\text{curv} c^2 \).  With this last definition and using the
fact that the Laplace-Beltrami operator applied to a scalar field \( \psi \) is
given by  \citep[see e.g.][]{daufields}:

\begin{equation}
  \Delta \psi = \frac{ 1 }{ \sqrt{-g} } \partial_\mu \left( 
    \sqrt{ -g } \, g^{\mu\nu} \partial_\nu \psi \right),
\label{laplace-beltrami}
\end{equation}

\noindent then 

\begin{equation}
  \rho_{\text{curv}} = \frac{ c^2 }{ 8\pi G F_R } \left[ \frac{ 1 }{ 2 }
    \left( R F_R -F \right) - \frac{ 3H }{ c^2 } \frac{ \mathrm{d} F_R 
    }{ \mathrm{d} t }\right],
\label{eq15}
\end{equation}

\noindent where \( H := \dot{a}(t)/ a(t) \) represents Hubble's constant.

  With the above definitions and using the \( 00 \) component of the
field's equations~\eqref{eq12} and the relation \citep[cf.][]{dalarsson}:

\begin{equation}
  R = - \frac{ 6 }{ c^2 } \left[ \frac{ \ddot{a} }{ a } +
    \left( \frac{ \dot{a} }{ a } \right)^2 + \frac{ \kappa c^2 }{ a^2 }
    \right],
\label{eq16}
\end{equation}

\noindent between Ricci's scalar and the derivatives of the scale factor
for a FLRW universe, then the dynamical  Friedman's-like equation for a
dust flat universe is:

\begin{equation}
  H^2 = \frac{ 8\pi G }{ 3 } \left[ \left( 1 + \frac{ c^4 }{ 8\pi G }
    F_T \right) \frac{ \rho }{ F_R } + \rho_{ \text{curv} } \right].
\label{friedmann}
\end{equation}

  The energy conservation equation is given by the null divergence
of the energy-momentum tensor:

\begin{equation}
  \left( \frac{ 8\pi G }{ c^4 } + F_T \right) \left( \dot{\rho} + 
    3 H \rho \right) = - \rho \frac{\mathrm{d}F_T}{\mathrm{d}t}.
\label{continuity}
\end{equation}

  For completeness, we write down the correspondent generalisation of 
Raychadhuri's equation for a dust flat universe:

\begin{equation}
  2 \frac{ \ddot{a} }{a} + H^2 = - \frac{ 8 \pi G p_\text{curv} }{c^2},
\label{raychadhuri}
\end{equation}

\noindent where the ``curvature-pressure'' 

\begin{equation}
  p_\text{curv} := \omega c^2 \rho_\text{curv},
\label{pcurv}
\end{equation}
  
\noindent and 

\begin{equation}
  w =  \frac{ c^2 \left( F - R F_R \right) / 2 + 
    \mathrm{d}^2 F_R / \mathrm{d} t^2  +  3 H 
    \mathrm{d} F_R / \mathrm{d} t }{ c^2  \left( R F_R 
    - F \right) / 2 -  3 H  \mathrm{ d } F_R /
    \mathrm{d} t }. 
\label{omega}
\end{equation}

   On the other hand, note that the mass \( M \) that appears on the
length \( L_M \) must be the causally connected mass at a certain cosmic
time \( t \), since particles beyond Hubble's (or particle) horizon with
respect to a given fundamental observer do not have any gravitational
influence on him.  At any particular cosmic epoch, this Hubble mass
satisfies the spherically symmetric condition implicit in
equation~\eqref{eq041} and so,

\begin{equation}
  M = 4 \pi \int_{0}^{r_\text{H}}{ \rho \, r^2 \, \mathrm{d}r }
    = \frac{4}{3}\pi \rho \frac{ c^3 }{ H^3 },
\label{mass}
\end{equation}

\noindent where 

\begin{equation}
  r_\text{H} := \frac{ c }{ H(t) },
\label{hubble-radius}
\end{equation}

\noindent is the Hubble radius or the distance of causal contact at a
particular cosmic epoch \citep{galaxy-formation}.  In this respect the
mass \( M \) is measured from the point of view of any given fundamental
observer at a particular cosmic time \( t \) and so, it does not depend
on which system of reference (or coordinates) is measured.  As such,
the mass \( M \) represents an invariant scalar quantity.  From this
last relation it follows that the length~\eqref{eqlm} is given by:

\begin{equation}
  L_M = \zeta \frac{ \left( \frac{4}{3} \pi c^3 G \right)^{3/4} }{ 
    c \, \mathsf{a}_0^{1/4} } \frac{ \rho^{3/4} }{ H^{9/4} }, 
\label{eq19}
\end{equation}

\noindent and so, by using relation~\eqref{fchi-power} and the standard
power-law assumptions:

\begin{equation}
  a(t) = a(t_0) \left( \frac{ t }{ t_0 } \right)^{\alpha}, \qquad 
  \rho(t) = \rho_0 \left( \frac{ a }{ a(t_0) } \right)^{\beta}.
\label{power-laws}
\end{equation}

\noindent for the unknown constant powers \( \alpha \) and \( \beta \),
it follows that:

\begin{gather}
  \frac{ \mathrm{d}F_R }{ \mathrm{d} t } = b (b-1) R^{b-1} 
    L_M^{2(b-1)} H \left[\frac{ j - q - 2 }{ 1 - q }+ \frac{ 3 }{ 2 }
    \left( \beta + \frac{ 3 }{ \alpha } \right) \right],
    					\label{eq20}\\
  \frac{ \mathrm{d}F_T }{ \mathrm{d}t } = \frac{ 3 }{ 2 }( b - 1 ) 
    \frac{ R^b L_M^{ 2b-2 } }{ \rho c^2 },
\label{derivadas_frft}
\end{gather}

\noindent where

\begin{equation}
  q(t):= -\frac{ 1 }{ a }\frac{ \mathrm{d}^2 a }{ \mathrm{d} t^2 } H^{-2}, 
    \qquad \text{ and } \qquad 
    j := \frac{ 1 }{ a } \frac{ \mathrm{d}^3 a }{ \mathrm{d} t^3 } H^{-3},
\label{j}
\end{equation}

\noindent are the deceleration parameter and the jerk respectively.

  With these and the value of \( L_M \) from equation~\eqref{eq19}, 
the curvature density~\eqref{eq15} is given by:

\begin{equation}
  \rho_\text{curv} = \frac{ 3 H^2 }{ 8 \pi G } ( b - 1 ) 
    \left[ \left( 1 - q \right) - \frac{ j - q - 2}{ 1 - q } -
    \frac{ 3 }{ 2 } \left( \beta + \frac{ 3 }{ \alpha } \right) 
    \right].
\label{dens_curv1}
\end{equation}

\noindent Substitution of the previous relations on Friedmann's
equation~\eqref{friedmann} gives:

\begin{equation}
  H^2 = \frac{ 8 \pi G \rho }{ 3 \, Z \, F_R },
\label{eq21}
\end{equation}

\noindent where

\begin{equation}
  Z := 1 + \left( b - 1 \right) \left[ \frac{ j - q - 2 }{ 1 - q } - 
    \frac{ 4 \left( 1 - q \right) }{ b } + \frac{ 3 }{ 2 } \left(
    \beta + \frac{ 3 }{ \alpha } \right) \right].
\label{Z}
\end{equation}

\noindent is a dimensionless function.

  An important result can be obtained evaluating equation~\eqref{eq21}
at the present epoch, yielding:

\begin{equation}
  \mathsf{a}_0 = \left[ \frac{ 9 }{ 4 } \zeta^4 \left( 1 - q_0 \right)^2
    \left( b Z_0 \right)^{ 2 / \left( b - 1 \right) }
    \left( \Omega_\text{matt}^{(0)} \right)^{ \left( 3 b - 5 \right) /
    \left( b - 1 \right) } \right] c \, H_0,
\label{a0ch0}
\end{equation}

\noindent where the density parameter \( \Omega_\text{matt}^{ (0) } \) 
at the present epoch has been defined by it's usual relation:

\begin{equation}
  \Omega_\text{matt}^{ (0) } := \frac{ 3 H^2 \rho }{ 8 \pi G }.
\label{omegazero}
\end{equation}

  In other words, the value of Milgrom's acceleration constant \(
\mathsf{a}_0 \)  at the current cosmic epoch is such that 

\begin{equation}
  \mathsf{a}_0 \approx c \times H_0.
\label{eq23}
\end{equation}

\noindent The numerical coincidence between the value of Milgrom's
acceleration constant \( \mathsf{a}_0 \) and the  multiplication of the
speed of light \( c \) by the current value of Hubble's constant \(
H_0 \) has been noted since the early development of MOND \citep[see
e.g.][and references therein]{famaey11}.  Note that equation~\eqref{eq23}
means that this coincidence relation occurs at approximately the present
cosmic epoch in complete agreement with the results by \citet{bernal11a}
where it is shown that \( \mathsf{a}_0 \) shows no cosmological evolution
and hence it can be postulated as a fundamental constant of nature.

  For the power law~\eqref{fchi-power} and the
assumptions made above, it follows that the energy conservation
equation~\eqref{continuity} is given by:

\begin{equation}
  \left( \dot{\rho} + 3 H \rho \right) + \frac{ c^2 }{ 8 \pi G }
    \left( A \frac{ \dot{\rho} }{ \rho} +B \, H \right) R^b \, 
    L_M^{2(b-1)}=0,
\label{continuity-simplified}
\end{equation}

\noindent where:

\begin{gather*}
  A := \frac{ 9 }{ 4 }\left( b - 1 \right)^2,
  				\\
  B := \frac{ 9 }{ 2 } \frac{ b - 1 }{ b } + \frac{ 27 }{ 4 } \frac{
    \left( b - 1 \right)^2 }{ \alpha } + \frac{ 3 }{ 2 } 
    \frac{ b \left( b - 1 \right) \left( j - q - 2 \right) }{ 1 - q }.
\end{gather*}

\noindent Direct substitution of the density power law~\eqref{power-laws} into
relation~\eqref{continuity-simplified} gives a constraint equation between \(
\alpha \), \( \beta \) and \( b \):

\begin{equation}
  \beta = \frac{ 1 }{ \alpha } \left( \frac{ 9 - 5 b }{ 3 b - 5 }\right).
\label{constraint}
\end{equation}

\begin{figure}
\begin{center}
  \includegraphics[scale=0.70]{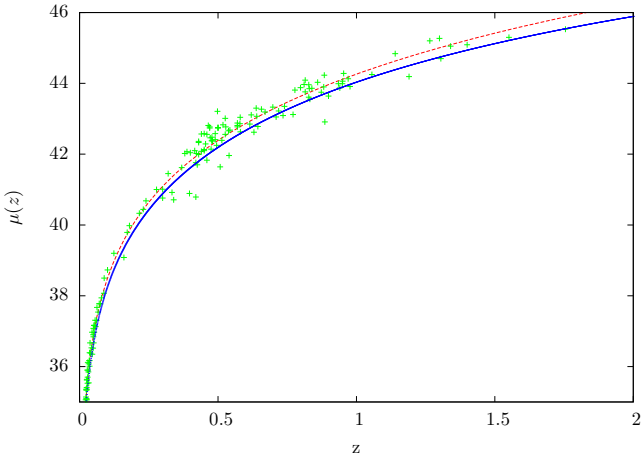}
  \caption[Best fit to SNIa data]{
	  Redshift magnitude plot for SNIa showing the  distance modulus
	  \( \mu \) as a function of the redshift \( z \) for SNIa as
	  presented by \citet{riess04}.  The dotted red line shows the best
	  fit to the data with the \( f(\chi) \) gravity theory applied to
	  a flat dust FLRW universe (see text) with no dark components.	
	  The continuous blue line represents the best fit according to the
	  standard concordance dust \( \Lambda \)CDM model.
          }
\label{fig02}
\end{center}
\end{figure}

  Let us now proceed to fix the so far unknown parameters of the theory \(
\alpha \), \( \beta \) and \( b \).   To do so, we need reliable
observational data and as such, we use the redshift-magnitude SNIa data 
obtained by \citet{riess04} and the following well known 
standard cosmological relations~\citep[see e.g.][]{galaxy-formation}:

\begin{gather}
  1 + z = a(t_0) / a(t),
  				\label{eq24} \\
  \mu\left(z\right) = 5 \log_{10}\left[ H_0 \, d_L \left( z \right )\right] - 
    5 \log_{10} h + 42.38,
  				\label{eq25} \\
  d_\text{L}\left(z\right) = \left( 1 + z \right) \int_{0}^{z} 
    \frac{ c }{ H \left( z \right)}\,\ \mathrm{d}z,
 				\label{eq26}
\end{gather}

\noindent for the cosmological redshift \( z \), the distance modulus \(
\mu \), the luminosity distance \( d_L \) and where the normalised Hubble
constant \( h \) at the present epoch is given by \( h := H_0 / \left(
100 \, \mathrm{km} \, \mathrm{ s }^{-1}  / \, \mathrm{Mpc} \right) \).
Also, from equation~\eqref{power-laws} it follows that

\begin{equation}
  H(a) = H_0\left(\frac{a}{a(t_0)}\right)^{-1/\alpha} = H_0(1+z)^{1/\alpha},
\label{solution_H}
\end{equation}
  
\noindent and the substitution of this into equation~\eqref{eq26} gives the
distance modulus \( d_L \) as a function of the redshift \( z \).  This
means that the redshift magnitude relation~\eqref{eq25} is a function that
depends on the values of the current Hubble constant \( H_0 \) and the
value of \( \alpha \).  Figure~\ref{fig02} shows the best fit to the
redshift magnitude relation of SNIa observed by \citet{riess04}, 
yielding \( \alpha=1.359 \pm 0.139 \) and  \( h = 0.64 \pm 0.009 \).  The
best fit presented on the figure was obtained using the Marquardt-Levenberg
fit provided by gnuplot (http://www.gnuplot.info) for non-linear functions.
These values do not provide the whole description of the problem,
since \( \beta \) and \( b \) are still unknown. However, according to the
constraint equation~\eqref{constraint} only one of them is needed in order
to know the other once \( \alpha \) is known.

  The parameter \( \beta \) can be found from conservation of mass
arguments, since the total mass of the universe \( M_\text{tot} =
4 \pi \int_0^{R_\text{max}}{\rho \, r^2 \, a^3 \, \mathrm{d}r } = 
\text{const.}\),  where the upper limit of the integral 
is the radius of the whole universe.  Since \( a(t) \) and \(
\rho(t) \) are time dependent functions, the only way the mass of the
universe is conserved is by requiring \( a^3 \, \rho  = \text{const.}\) and
so, \( \beta = 3 \).  This argument is exactly the one used in
standard cosmology when dealing with a dust FLRW universe
\citep[see e.g.][]{galaxy-formation}.  Using this value of \( \beta \) and
the one already found for \( \alpha \), it follows that \( b = 1.57 \pm 0.56
\), which is within the expected value of \( b = 3/2 \)  discussed in
section~\ref{relativistic-extension}.

  For completeness, we write down a few of the cosmographycal parameters
obtained by this \( f(\chi) \) gravity applied to the universe:

\begin{equation}
   \begin{gathered}
   h = 0.64 \pm 0.009,  \quad q_0 = -0.2642 \pm 0.075, \\
   j_0 = -0.1246 \pm 0.004.
   \end{gathered}
\label{eq27}
\end{equation}


\section{Discussion}
\label{discussion}

  As explained by \citet{carranza12}, the obtained value \( b \approx
3/2 \) is a completely expected result due to the following arguments.
As explained in section~\ref{extended-newtonian}, a gravitational
system for which its characteristic size \( r \) is such that \( x :=
l_M / r \lesssim 1 \) is in the MONDian gravity regime.  For the case
of the universe, \( x \sim \text{a few} \) and as such if not totally
in the MONDian regime of gravity, then it is far away from the regime
of Newtonian gravity.  The relativistic version of this means that the
universe is close to the regime for which \( f( \chi ) = \chi^{3/2}
\) and so \( b = 3/2 \).  This is a very important result since seen
in this way, the accelerated expansion of the universe is due to an
extended gravity theory deviating from general relativity.  It is quite
interesting to note that the function \( f(\chi) = \chi^{3/2} \) which at
its non-relativistic limit is capable of predicting the correct dynamical
behaviour of many astrophysical phenomena, is also able to explain the
behaviour of the current accelerated expansion of the universe.

  Seen in this way, the behaviour of gravity towards the past (for
sufficiently large redshifts \( z \)) will differ from \( f(\chi) =
\chi^{3/2} \) and eventually converge to \( f(\chi) = \chi \), i.e. the
gravitational regime of gravity is general relativity for sufficiently
large redshifts. A very detailed investigation into this needs to be done
at different levels in order to be coherent many different cosmological
observations \citep[see e.g.][]{longair11}.  This in turn can serve to
calibrate the index \( p \) of the transfer function \( f(\chi) \) as
presented in equation~\eqref{eq032}, which has a very soft transition when
\( p = -1 \), i.e.,

\begin{equation}
  f(\chi) = \frac{ \chi^{3/2} }{ 1 + \chi^{3/2} },
\label{eq28}
\end{equation}

\noindent  and also has a very sharp transition when \( p \rightarrow
\infty \), with the step function:

\begin{equation}
  f(\chi) = 
  \begin{cases}
    \chi^{3/2}, \quad \text{for }  0 \leq \chi \leq 1, \\
    \chi, \qquad \ \text{for } \chi \geq 1.  
    \end{cases}
\label{eq29}
\end{equation}

\noindent  In this respect, perhaps something close to a sharp
transition~\eqref{eq29} will be observed since, as mentioned in
section~\ref{extended-newtonian}, at the non-relativistic level different
astrophysical observations show a sharp transition from the Newtonian
to the MONDian regimes.  This sort of decision has to be taken with care
and such a full description requires to analyse in full detail the whole
Friedmann-like equations:

\begin{gather}
    \begin{split}
      \left( \frac{ 8 \pi G }{ c^4 } + F_T \right) & \left( \dot{ \rho } + 
        3 H \rho +\frac{ 3 H p }{ c^2 } \right)  = \\
        & - \rho \frac{ 
        \mathrm{d} F_T }{ \mathrm{d}t } + \frac{ 1 }{ c^2 } \left( p 
        \frac{ \mathrm{d} F_T}{ \mathrm{d} t } + F_T \frac{ \mathrm{d} p 
        }{ \mathrm{d} t } \right),
    \end{split}
					\label{eq30} \\
    H^2 = \frac{8 \pi G }{ 3 } \left[ \left( 1+ \frac{ c^4 F_T }{ 
      8 \pi G } \right) \frac{ \rho }{F_R} + \rho_{\text{curv}} \right] - 
      \frac{ \kappa c^2 }{ a^2 },
      					\label{eq31} \\
    2 \frac{ \ddot{a} }{ a } + H^2 + \frac{ \kappa c^2 }{ a^2 } = 
      - \frac{ 8 \pi G p }{ c^2 F_R } - \frac{ 2 p c^2 F_T }{ F_R } - 
      \frac{ 8 \pi G p_\text{curv} }{ c^2 }.
      					\label{eq32} 
\end{gather}

\noindent These equations are directly obtained from taking
the null covariant divergence of the energy momentum tensor, the \(
00 \) component of the field equations~\eqref{eq12} and the density \(
\rho \) contains all species of matter and/or radiation.  The curvature
density \( \rho_\text{curv} \) and the curvature pressure \( p_\text{curv}
\) are related to one another by relation~\eqref{pcurv} with \( \omega
\) given by equation~\eqref{omega}.

  It is quite remarkable that a metric extended theory of gravity is able
to reproduce phenomena from mass and length scales associated to 
the solar system up to cosmological scales.  There are many more
astrophysical challenges that this theory needs to address, in particular
with respect to lensing at different scales and the dynamics associated to
galaxy clusters.  These will be addressed elsewhere.

\section{Acknowledgements}
\label{acknowledgements}

  This work was supported by a DGAPA-UNAM grant (PAPIIT IN116210-3) and
CONACyT 26344.  The author acknowledges fruitful discussions at different
stages  with Tula Bernal, Diego Carranza, Salvatore Capozziello, 
Rituparno Goswami, Xavier Hernandez, Juan Carlos Hidalgo and Luis Torres.

\bibliography{a0isH0}

\end{document}